\documentclass[a4paper,twocolumn]{article}

\usepackage{articleExtension}

\begin{document}
\title{SPIRAL2 Cryomodules Models: a Gateway to Process Control and Machine Learning}
\author{A. Vassal, A. Ghribi, F. Millet, F. Bonne, P. Bonnay, P.-E. Bernaudin
\thanks{Submitted on 27/02/2121. This work was supported in part by the Caen-Normandie university.}
\thanks{A. Vassal was with the DSBT and GANIL laboratories, he is now with the Corys compagny. (e-mail: vassal.adrien.pro@gmail.com).}}

\maketitle

\begin{abstract}
\noindent From simple physical systems to full production lines, numerical models could be used to minimize downtime and to optimize performances. In this article, the system of interest is the SPIRAL2 (Système de Production d'Ions RAdioactifs en Ligne de 2e génération) particles accelerator cryogenic system. This paper illustrates three totally different applications based on a SPIRAL2 cryostat model: optimal controller synthesis, virtual sensor synthesis and anomaly detection. The tow firsts applications have been deployed on the system. Experimental results are used to illustrate the benefits of such applications.   
\end{abstract}

%
\section{Introduction}
SPIRAL2 (Système de Production d'Ions RAdioactifs en Ligne de 2e génération) is a heavy ions accelerator located in Caen (France). Its main part is a linear accelerator (LINAC) \cite{Ferdinand2008} composed of 26 bulk niobium RF resonators which accelerate charged particles by the mean of electromagnetic fields \cite{Padamsee2009}. To be operated, those resonators need to be maintained in a supercondicting state. As the niobium transition temperature is 9.2~$K$, a cryogenic system is required. Resonators are coupled with an RF system, as well as vacuum and cryogenic components. The assembly of these subsystems forms a cryomodule.\\
The cooling power is provided by a cryoplant with a maximal capacity of 1300~$W$ at 4.2~$K$. A cold box coupled with a 5000~$L$ helium dewar provides liquid helium to all the cryomodules through a cryodistribution. Inside of the cryomodules, the liquid helium evaporates to extract heat from the resonator and gaseous helium is returned to the coldbox. More details on the cryogenic system can be found in \cite{Ghribi2017,Ghribi2017_2}.\\
As a perturbation in the cooling system might lead to a shutdown of the accelerator, it is mandatory to develop a high reliability operating system. To achieve this, modeling tools are developed to improve the control robustness, to predict highly valuable information and to realize fault detection.\\
The present paper mainly focuses on the cryomodules and not on the overall cryogenic system.\\
The first section is dedicated to the modeling of the cryomodules. In the second section, an optimized control law is proposed. The third section details the synthesis of a virtual sensor used to predict unmeasured parameters. The last section is dedicated to fault detection by the mean of machine learning.

\section{Modeling of the cryomodules}

	\subsection{Description}

There are two types of cryomodules, namely a type-A and type-B. The main difference between them lays in the fact that the type-A contains one resonating cavity whereas the type-B contains two of them. For more details concerning the cryomodules design and performances, please refer to \cite{Bernaudin2004,Olry2006}.\\

In terms of thermodynamics, both types of cryomodule undergoes different heat loads. First, the static heat loads induced by the heat transfer between the cold parts and their surrounding. Second, the dynamic heat loads due to the RF resistive losses in the resonating cavity. Heat load amplitudes are different for the two types of cryomodules. For the 4.4~$K$ bath, those characteristics as well as the volumes are given in Table \ref{tab:CryomoduleComparison}.
\begin{table}[!htb]
	\centering
	\begin{tabular}{|c|c|c|}
	\hline
	Characteristics		   & Type-A      & Type-B 		\\
	\hline
	Helium bath volume [L] & 20.5        & 91.5 		\\ 
	\hline
	Static heat load [W]   & 3.5$\pm$1.4 & 12.5$\pm$1.8 \\
	\hline
	RF heat load [W]       & 5.8$\pm$2.2 & 12.1$\pm$2.6 \\
	\hline
	\end{tabular}
	\caption{Main differences between type-A and type-B cryogenic systems. Heat loads are measured on helium phase separator.}
	\label{tab:CryomoduleComparison}
\end{table}
\begin{figure*}[!t]
	\centering
	\psfragfig*[width=0.7\linewidth]{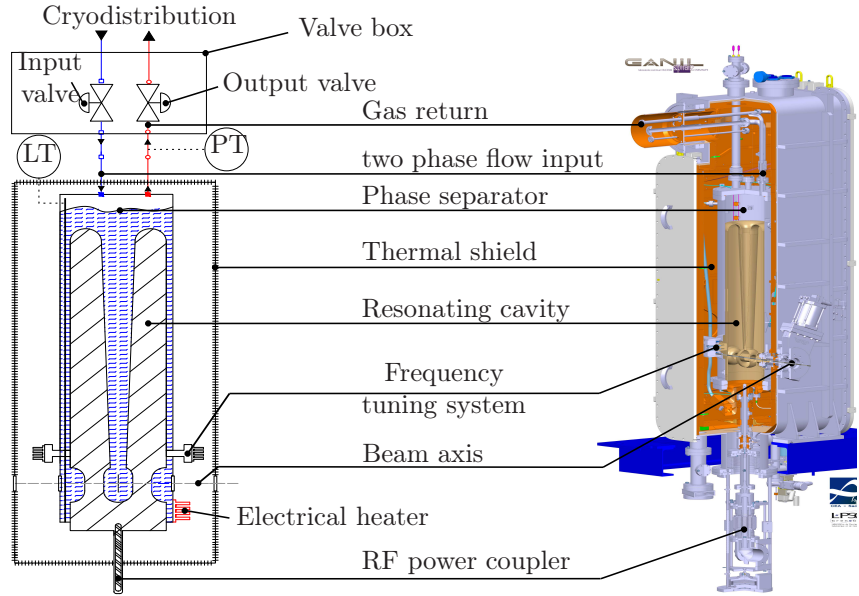}{
	\psfrag{tag0}{Cryodistribution}
	\psfrag{tag1}{Output}
	\psfrag{tag2}{\begin{tabular}{@{}c@{}}
   						Input\\
   						valve\\
				  \end{tabular}}
	\psfrag{tag3}{LT}
	\psfrag{tag4}{Output valve}
	\psfrag{tag5}{PT}
	\psfrag{tag6}{Gas return}
	\psfrag{tag7}{two phase flow input}
	\psfrag{tag8}{Thermal shield}
	\psfrag{tag9}{Resonating cavity}
	\psfrag{tag10}{\begin{tabular}{@{}c@{}}
   						Frequency\\
   						tuning system\\
					  \end{tabular}}
	\psfrag{tag11}{RF power coupler}
	\psfrag{tag12}{Electrical heater}
	\psfrag{tag13}{Valve box}
	\psfrag{tag14}{Phase separator}
	\psfrag{tag15}{Beam axis	}
	}
	\caption[Cut view and 3D view of a Type-A cryomodule]{Representation of a type-A cryomodule. On the left, a simplified scheme. On the right a 3D cut view. LT and PT are respectively level transmitter and pressure transmitter.}
	\label{fig:Cryomodule_3Dview_PIDview}
\end{figure*}

The cryogenic system is in charge of keeping the cavities under the critical temperature. It is composed of three main elements as shown in \ref{fig:Cryomodule_3Dview_PIDview}: a phase separator filled with liquid helium at 4.4 K at 1200 mbar, a thermal shielding that surrounds the phase separator, and a valve box containing all the valves used to control cryogenic operating conditions. As the phase separator is the most critical element of the cryogenic system, we will only focus on that element and its associated valves. The Figure \ref{fig:Cryomodule_3Dview_PIDview} presents a simplified scheme with the subsystems of interest. 
The phase separator is fed with liquid helium through the input valve which is used to regulate the level of liquid. Due to thermal heat load, liquid helium evaporates and is returned to the cold box. In that process, gas goes through the output valve which is used to regulate the pressure within the phase separator.\\
Both valves and phase separator have been modeled. Valves equations are the ones given in the standard ISA \cite{ISA2012}, whereas phase separator dynamics are describes through energy and mass balance. The equations have been implemented in the Simcryogenics library \cite{Bonne2020}, a modelling tool used to simulated and optimize cryogenic systems. Helium properties are extracted from tabulated data using HEPAK package \textsuperscript{\textregistered}. As those equations have been extensively described in \cite{Vassal2019,Vassal2018_1,Vassal2018_2}, they will not be discussed in this article. Rather than that, the comparison between experimental and simulation results is emphasized.

	\subsection{Model vs Data}

\begin{figure}[!b]
	\centering
	\psfragfig*[width=\linewidth]{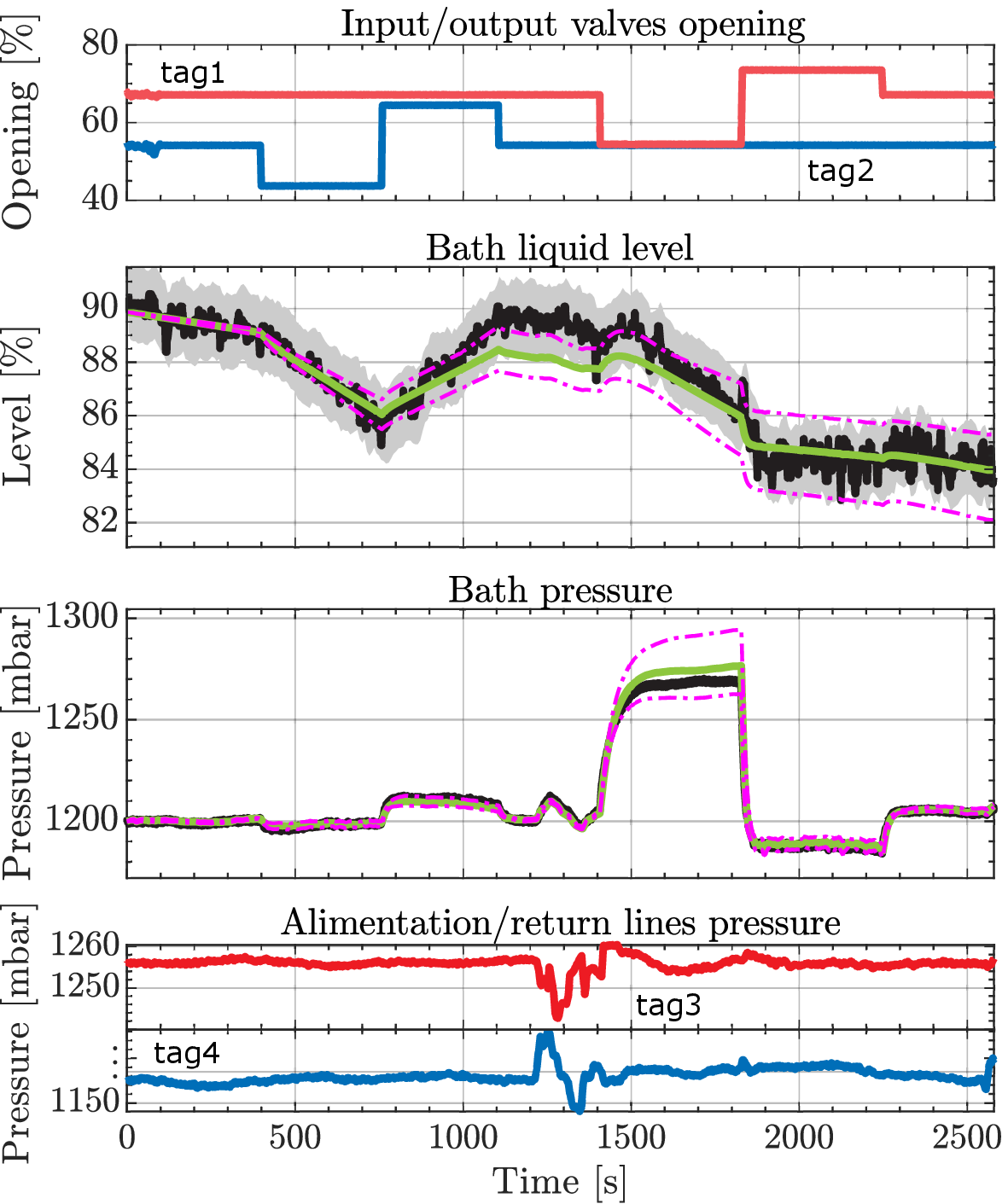}{
	\psfrag{tag1}{\textcolor[rgb]{0.9294,0.1059,0.1373}{output}}
	\psfrag{tag2}{\textcolor[rgb]{0,0.4314,0.7216}{input}}
	\psfrag{tag3}{\textcolor[rgb]{0.9294,0.1059,0.1373}{alimentation}}
	\psfrag{tag4}{\textcolor[rgb]{0,0.4314,0.7216}{return}}
	}
	\caption{Model vs measurement for the first type-A cryomodule. Measurement and associated uncertainty are respectively black line and grey background. Simulation value and uncertainty are respectively green line and magenta dash-dotted line.}
	\label{fig:CMA_model_vs_measure}
\end{figure}

The simulation results for both cryomodule types have been compared to experimental data. For each of the cryomodules, an operating scenario has been performed starting from stable operating conditions. This scenario is a series of steps applied on the input and output valves opening command. The same values have been applied to the model and to the real process in an open loop manner. The comparison obtained for the cryomodule 1 (the first one on the line considering the beam direction) is shown in Figure \ref{fig:CMA_model_vs_measure}.

\begin{figure}[!htb]
	\centering
	\psfragfig*[width=\linewidth]{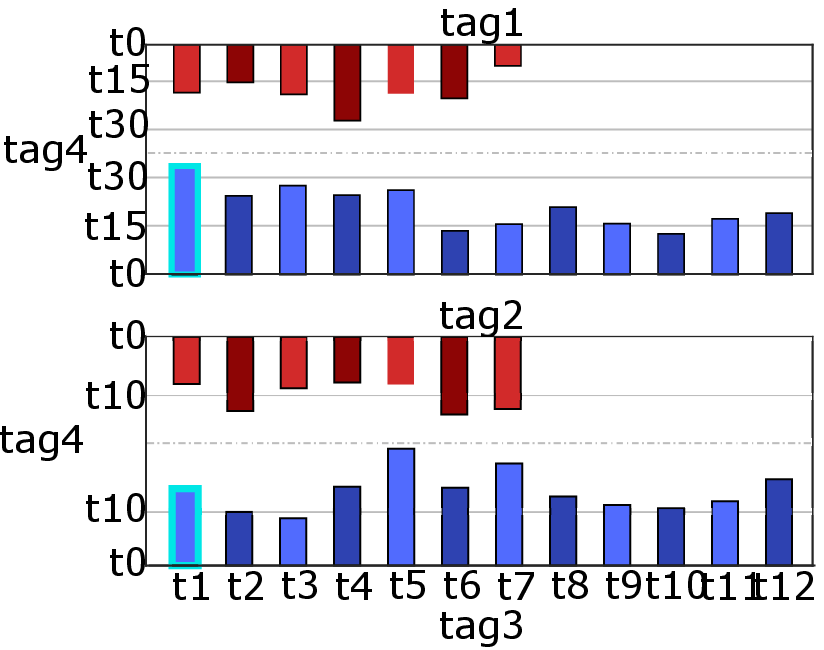}{
	\psfrag{tag1}[cc][cc][1][0]{$Cr$ evaluates on level}
	\psfrag{tag2}[cc][cc][1][0]{$Cr$ evaluates on pressure}
	\psfrag{tag3}[cc][cc][1][0]{Cryomodule number}
	\psfrag{tag4}[cc][cc][1][90]{$Cr$}
	\psfrag{t0}{0}
	\psfrag{t1}{1}
	\psfrag{t2}{2}
	\psfrag{t3}{3}
	\psfrag{t4}{4}
	\psfrag{t5}{5}
	\psfrag{t6}{6}
	\psfrag{t7}{7}
	\psfrag{t8}{8}
	\psfrag{t9}{9}
	\psfrag{t10}{10}
	\psfrag{t11}{11}
	\psfrag{t12}{12}
	\psfrag{t15}{15}
	\psfrag{t30}{30}
	}
	\caption{Evaluation of the criteria $Cr$ on each cryomodule. The type-A and type-B cryomodule are respectively plotted in blue and red. The blue highlighted element correspond to the same cryomodule than the one in Figure \ref{fig:CMA_model_vs_measure}.}  
	\label{fig:CRComparisonResult}
\end{figure}

The comparison shows a good agreement between experimental and simulated data for both level and pressure dynamics. It's worth mentioning that the uncertainty of the modeled level increases with time as the level is an integrator system. Furthermore, the high uncertainty on the pressure at time t=1500~$s$ is mostly due to the valve position uncertainty: an error of $\pm$~1\% on valve position could lead to a pressure uncertainty up to 10~$mbar$. Finally, the pressure peak occurring at t=1200~$s$ is due to a pressure oscillation in the cryodistribution (i.e. the inlet boundary of the model).\\
Similar results have been observed on the other cryomodules, the following criteria have been calculated for each comparison:
\begin{equation}
	Cr = \frac{100 \cdot \int_{t_{init}}^{t_{final}} \left(V_{mes}(t)-V_{sim}(t)\right)^2 dt}{V_{moy}^{init} \cdot t_{total}}
\end{equation}	
Where $Cr$ is the criteria representing the integral of the error between measured and simulated data. $t_{init}$, $t_{final}$ and $t_{total}$ are respectively the initial time, the final time and the overall time of the scenario, whereas $V_{mes}(t)$ and $V_{sim}(t)$ are respectively the measured and simulated values. Finally, $V_{moy}^{init}$ designate the mean value at the beginning of the scenario.
The division by the overall duration and the mean value makes this criteria adimensional. So, it is possible to compare multiple scenarios with different durations and operating conditions.\\
The criteria values obtained for the scenario shown in Figure \ref{fig:CMA_model_vs_measure} but applied to all the cryomodules are plotted in Figure \ref{fig:CRComparisonResult}.

As one can see, errors are all contained in a narrow range. It means that similar comparison results than the ones shown in Figure \ref{fig:CMA_model_vs_measure} have been observed for each cryomodule. It also shows that the same model could be used in a generic way (i.e. for different cryomodules) without any precision loss.
From the author point of view, the prediction capacity of the model is sufficient to be used in further applications. Three examples of these applications will be given in the following sections.

\section{Optimal controller synthesis}

	\subsection{Problem overview}
			
	Cryogenic system control loops are critical items that can affect the overall accelerator. Two requirements are to be met in order to allow the nominal operation of the RF cavity.\\
The first is to ensure that the cavity temperature remains below its critical temperature. Otherwise, the cavity could quench\footnote{fast transition between superconducting state and normal conducting state that can lead, in the worst case, to irreversible mechanical damages}. To do so, the cavity is submerged in a liquid helim bath and the level of liquid helium is regulated through a PID (Proportional Integrator Derivative) controller acting on the input valve (see Figure \ref{fig:Cryomodule_3Dview_PIDview}). The goal is to maintain a level at $90\% \pm 5\%$ which is high enough to maintain the overall cavity fully submerged with a comfortable operating margin.\\
The second is to ensure that the shape of the cavity doesn't change as the performances of the resonator are intrinsically linked to the cavity shape. This could be seen in the expression of its unloaded quality factor:
\begin{equation}
	Q_{f0} = \frac{G}{Rs} \label{eq:UnloadedQualityFactor}
\end{equation}
Where $Q_{f0}$ is the unloaded quality factor, $Rs$ the surface resistance and $G$ the geometric factor that depends on the surface and the volume of the cavity. As the cavity is submerged in liquid a helium bath, any pressure variation ($\Delta P$) in the separator will induce a mechanical force on the cavity walls that slightly deforms the cavity. This results in a variation of the geometric factor that can lead to a drop of the cavity quality factor hence significantly reducing the cavity acceleration gradient.\\
Considering the bandwidth of the cavity and its associated RF system, a pressure variation limit of $\Delta P = \pm 5\ mbar$ has been set up for the SPIRAL2 cryomodules. It is worth mentioning that the nominal pressure of the helium bath is $1\ 200\ mbar$, which mean that a $\Delta P$ of $\pm 5\ mbar$ represents a tolerance of $\pm 0.41 \%$. Both level and pressure are regulated through PID controllers. Although the PID performance is enough to achieve the level requirement, it isn't the case for pressure requirement. Even using a state of the art \cite{Matlab2019} PID (Proportional, Integral Derivative) tuning tool, we were not able to maintain the pressure variation within a range of $\pm 5\ mbar$ for long periods of time (i.e more than a few hours). This is probably due to the fact that the two regulation loops are coupled: an action on the input valve influences the level and the pressure. Similar statement is also true for the output valve: an action on the output valve has an impact both on pressure and level.\\
As PID controllers are more suitable in the case of linear SISO \footnote{Single Input Single Output} system, an other control algorithm is necessary to achieve the project requirement.

	\subsection{Synthesis of a LQ regulator}

\begin{figure}[!t]
	\centering
	\includegraphics[width=\linewidth]{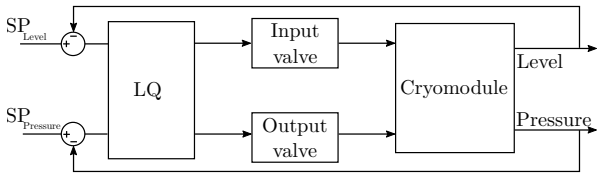}
	\caption[Block diagram of the synthesized LQ regulator]{Block diagram of the synthesized LQ regulator. SP designates the setpoint.}
	\label{fig:Cryomodule_LQ_block_diagram}
\end{figure}

A few parameters have to be considered while choosing the most suitable solution for the cryomodules control loops.\\
First, the cryomodule cryogenic system is a MIMO\footnote{Multiple Inputs Multiple Outputs} system with two valves as inputs and the level and pressure as outputs. As there is internal coupling between all inputs and outputs, a controller that can handle this coupling is mandatory.\\
Second, as the accelerator will be used for many years, it must be a solution proven on multiple systems with a full documentation.\\
Third, the controller has to be implemented in a dedicated PLC (Programmable Logical Controller) with limited amount of calculation capacity.\\
Considering those parameters, an LQ (Linear Quadratic) controller seems to be the best option. The block diagram of such controller applied to our system is given in Figure \ref{fig:Cryomodule_LQ_block_diagram}.\\
The mathematical development of this controller has already been described in \cite{Vassal2018_2, Vassal2019}. In this section we will only remind the main equations of the discrete LQ controller and focus on the latest experimental results.\\

The principle of a LQ controller is to synthesize a state feedback gain such that the command input is given by:
\begin{equation}
	u(k) = -K \cdot x(k) \label{eq:stateFeedback}
\end{equation}
Where $K$ is the state feedback gain and $x$ the state of the system. $K$ is calculated so that it minimizes the following quadratic cost:
\begin{equation}
	J = \sum_{i=k}^{\infty} x(i)^T \cdot Q \cdot x(i) + u(i)^T \cdot R \cdot u(i) \label{eq:QuadraticCost}
\end{equation}
Where $J$ is the cost, and $Q$ and $R$ are respectively state and input weights. As for gain and integral time for a PI controller, the goal is to tune the matrices $Q$ and $R$ to fulfill the process specifications. Details about the way to tune those gains are given in \cite{Vassal2019}.\\
The calculation of the state feedback gain $K$ requires the state-space model of the system which could be directly generated with the previously described model and a linearization algorithm such that the one described in \cite{Bonne2014}.\\
To allow a comparison between existing PID and the proposed regulation law, the LQ controller has been implemented on the existing PLC of each cryomodule. Even if they have a limited calculation capacity (a work memory of 192 Ko), it's more than enough for the proposed LQ controller which only requires around 30 multiplications/additions per sampling time. This is due to the fact that only the control law described in (\ref{eq:stateFeedback}) and its associated Luenberger observer \cite{Luenberger1971} have been implemented. The calculation of the state feedback gain $K$ that minimizes the cost (\ref{eq:QuadraticCost}) could be done offline using dedicated software like Octave or Matlab.

	\subsection{Experimental results}

\begin{figure}
	\centering
	\includegraphics[width=\linewidth]{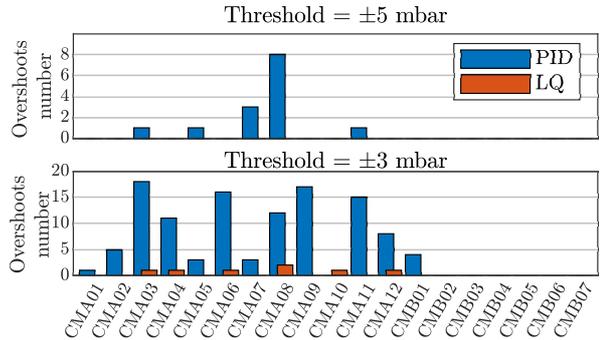}
	\caption{LQ vs PID pressure overshoot number for one night.}
	\label{fig:LQ_vs_PID_night}
\end{figure}

	Classical tests like setpoint variation and output disturbance test (using electrical heaters as disturbance sources) have been performed on both PID and LQ controllers. The results of those tests show that the LQ regulator is slightly better in term of robustness and speed than the PI controller. Nevertheless, our main interest is to see which controller is able to respect the level and pressure requirements during a long period of operation. To check this, the following experiment has been realized during two nights \footnote{Night is chosen to avoid daily operations that could induce comparison bias}. During the first night (12 hours), all cryomodules are regulated by PID controllers. During the second night they are regulated by LQ controllers. For both cases, the level requirement has been respected, but not the pressure requirement. To illustrate this, the number of times the pressure overshoots the threshold of $\pm 5\ mbar$ have been used as a metric to compare the two controllers. The result of the comparison is given on Figure \ref{fig:LQ_vs_PID_night}.
The LQ controller shows no pressure overshoot at all whereas PID controller shows multiple ones. However, the LQ controller could not perfectly dump pressure oscillations. To illustrate this, a counter of overshoot with a tighten pressure threshold of $\pm 3\ mbar$ (in comparison with the specification of $\pm 5\ mbar$) has also been plotted on Figure \ref{fig:LQ_vs_PID_night}.\\
Nevertheless, the results obtained with the LQ regulator are satisfying considering process requirements. As the algorithm has been deployed in the PLCs for the purpose of the test, it's already available for the current operation. This new control strategy is an important improvement that could reduce the accelerator downtime as one pressure overshoot may raises safety chains that shutdown the accelerator beam.

\section{Virtual sensor}
	
	\subsection{Problem overview}
		
\begin{figure*}[!ht]
	\centering
	\psfragfig*{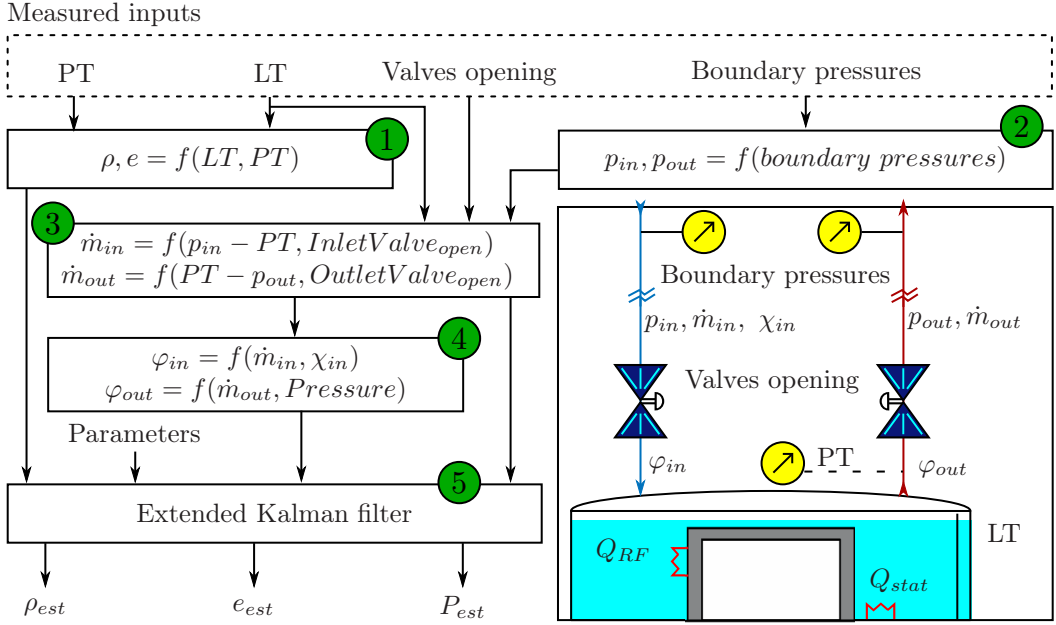}{
		\psfrag{tag0}{Measured inputs}
		\psfrag{tag1}[cc][cc][1][0]{PT}
		\psfrag{tag2}[cc][cc][1][0]{LT}
		\psfrag{tag3}[cc][cc][1][0]{Valves opening}
		\psfrag{tag4}[cc][cc][1][0]{Boundary pressures}	
		\psfrag{tag6}[cc][cc][1][0]{$\rho, e = f(LT,PT)$}		
		\psfrag{tag7}[cc][cc][1][0]{$p_{in},p_{out}=f(boundary\ pressures)$}		
		\psfrag{tag9}[cc][cc][1][0]{{\begin{tabular}{@{}c@{}}
   						$\dot m_{in}=f(p_{in} - PT, InletValve_{open})$\\
   						$\dot m_{out}=f(PT - p_{out}, OutletValve_{open})$
					  \end{tabular}}}
		\psfrag{tag10}[cc][cc][1][0]{{\begin{tabular}{@{}c@{}}
   						$\varphi_{in} = f(\dot m_{in}, \chi_{in})$\\
   						$\varphi_{out} = f(\dot m_{out}, Pressure)$
					  \end{tabular}}}
		\psfrag{tag11}[cc][cc][1][0]{Parameters}
		\psfrag{tag12}[cc][cc][1][0]{Extended Kalman filter}
		\psfrag{tag13}[cc][cc][1][0]{$\rho_{est}$}
		\psfrag{tag14}[cc][cc][1][0]{$e_{est}$}
		\psfrag{tag15}[cc][cc][1][0]{$P_{est}$}
		\psfrag{tag16}[cc][cc][1][0]{Boundary pressures}
		\psfrag{tag17}{$p_{in}, \dot m_{in},\ \chi_{in}$}
		\psfrag{tag18}{$p_{out}, \dot m_{out}$}
		\psfrag{tag19}[cc][cc][1][0]{Valves opening}
		\psfrag{tag21}{$\varphi_{in}$}
		\psfrag{tag22}{$\varphi_{out}$}
		\psfrag{tag23}[cc][cc][1][0]{$Q_{RF}$}
		\psfrag{tag24}[cc][cc][1][0]{$Q_{stat}$}
	}
	\caption{Block diagram of the virtual sensor and its associated model schematic view. $f$ designates different functions depending on the associated bloc number. $LT$ and $PT$ are respectively level and pressure transmitters. Numbers within green circles are explained in the text.}
	\label{fig:Cryomodule_Observer_block_diagram}
\end{figure*}

As the RF signal injected in the resonator is sinusoidal, it generates energy dissipation in the cavity walls called AC losses \cite{Wipf2003}. Those losses are considered as an indicator of the cavity state: an abrupt raise of those losses can indicates that a part of the cavity is no more in superconducting state. This could be the premises of a global quench of the cavity with potential irreversible mechanical damages. On another time scale, a slow increases of the dissipated AC losses can indicate a pollution of cavity with non-superconducting elements.\\
In the case of the SPIRAL2 project, there is no continuous measurement of these AC losses. Measurement can only be performed when the cavity is not in operation as the measurement method is intrusive \cite{Ding2010}. There is no operating solution in the case of SPIRAL2 that would allow us to perform such measurement online and without disturbing the process.\\

To solve this problem, we proposed a method to estimate these losses based on the phase separator model and an extended Kalman filter (EKF) \cite{Kalman1960}.

	\subsection{Synthesis of an extended Kalman filter}

From the phase separator point of view, the AC losses represent an external thermal heat load. The more AC losses, the more heat has to be extracted through the vaporization of liquid helium. Equations that link the AC losses to the thermal heat load are given in \cite{Vassal2018}. Knowing this, measuring the AC losses is equivalent to measuring the heat load extracted by the liquid helium bath.\\
Nevertheless, as for the AC losses, there is no continuous measurement of the heat load dissipated in each cavity in the SPIRAL2 cryogenic system. Discontinuous measurement can be made by measuring the evaporating rate of the liquid helium \cite{Adnan2020}, but once again it's an intrusive method that couldn't be realized during operation.\\
This is where the cryomodule model became very useful: using model and process measurement such as phase separator level and pressure as well as valves opening, it's possible to predict the current heat loads. Therefore, the idea is to synthesize an observer (also called a virtual sensor in that case) of the heat load.\\

An extended Kalman filter seems to be the best choice as it's designed to work with non linear process and has been successfully used in many applications \cite{McGee1985, Torres2020}. The process diagram of such an observer applied to our process is described in Figure \ref{fig:Cryomodule_Observer_block_diagram} where it's decomposed in elementary steps represented as number in green circles:
\begin{itemize}
	\item 1: calculate phase separator internal energy ($e$) and density ($\rho$) through property interpolation using bath pressure and liquid level
	\item 2: define model boundaries pressures ($p_{in}$ and $p_{out}$) using the  closest available pressure transmitters. This is equivalent to calculating a pressure drop between the closest sensors and the model boundaries based on the current mass flows and temperatures.
	\item 3: calculate input ($\dot m_{in}$) and output ($\dot m_{out}$) mass flows through valves considering valves pressure drop, valves opening and valves input quality ($\chi_{in}$). 
	\item 4: define phase separator input ($\varphi_{in}$) and output ($\varphi_{out}$) energy flux.
	\item 5: apply extended Kalman filter algorithm using model parameters (i.e. valves coefficients, bath volume and bath static heat loads) to estimate the heat loads dissipated in the phase separator.
\end{itemize}
In a nutshell, we use the difference between estimated values based on the model (i.e. $e_{est}$ and $\rho_{est}$) and values ($e$ and $\rho$) directly calculated from measurement (tabulated data in HEPACK), to correct the estimated heat load based on the model equations.\\
For the moment, the complete algorithm hasn't been deployed in a PLC. It has been tested on an external computer connected to the data acquisition system of the process. In that way, it was possible to get value from sensor but with a few seconds of delay.

	\subsection{Experimental results}
		
\begin{figure}
	\centering
	\includegraphics[width=\linewidth]{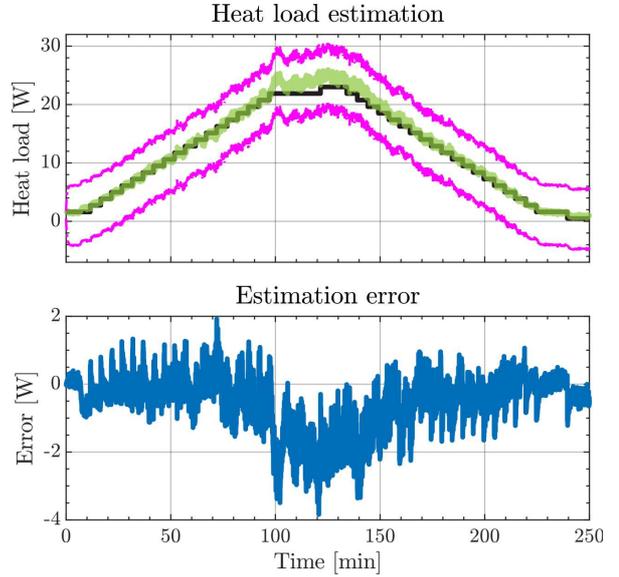}
	\caption{Estimated heat loads (green) and uncertainty (magenta) vs electrical heater setpoint (black).}
	\label{fig:CMA_estimation_vs_measure}
\end{figure}

To evaluate the estimation capacity of the extended Kalman filter, a reference was needed. The electrical heater located in the phase separator was the solution as its consumed current is measured. Considering that all the electrical power is transmitted to the phase separator as thermal heating, it's possible to compare the virtual sensor prediction to the injected electrical power. A test in which the electrical power is modified by steps of $1\ W$ is presented in Figure \ref{fig:CMA_estimation_vs_measure}.
The absolute average estimation error is equal to $0.7\ W$ which represents about $3\%$ of the maximal tested heat load. This mean that the estimation precision could reach $1\ W$ if the estimation is averaged on a sufficiently long timescale.\\
Nevertheless, for heat load above $20\ W$ the error estimation increases to $-2\ W$. This could be linked to an oversimplification of the dissipated heat load. In the experiment, we considered that the electrical power consumed by the heater is exactly equal to the dissipated heat. As the heater is installed on the external wall of the phase separator, it's possible that part of the electrical heating is dissipated by the thermal shield and not only in the helium bath. This would result in the observed bias between the real heat load and the one considered during the experiment.\\
This bias has not be measured yet but it could be quantified for different heat loads by measuring the liquid level decrease during a given time \cite{Adnan2020}. Then this value could be applied as a correction factor to improve the estimation.\\

However, the presented virtual sensor shows its capability to predict heat load in real time with a precision of few Watts. This is enough to detect anomalies during operation.\\
It is worth mentioning that the overall synthesis of this virtual sensor has been patented \cite{Vassal2018}. 

\section{Anomaly detection using machine learning}

	\subsection{Problem overview}

"Anomaly detection" is used to designate algorithms capable of identifying events or items differing from the majority of the events/items. For the case of plant monitoring these algorithms could be used to address the problem of continuous fault detection on process actuators or transmitters. This kind of algorithms are particularly suitable for large processes which contain thousands of actuators and transmitters, because it's almost impossible for a single operator to continuously check the functioning of each element within the process. For example, in the SPIRAL2 cryogenic system, there are more than 70 control valves and 300 transmitters.\\

In such case, we demonstrate the possibility to use the cryomodule model to perform actuator malfunction detection. To be more specific, we try to predict if the output valve of one cryomodule (see Figure \ref{fig:Cryomodule_3Dview_PIDview}) is undergoing deadband\footnote{A deadband is a range of input control that doesn't result in any output on valve position} problem. We use machine learning (ML) algorithms to predict the malfunction.\\

This section is decomposed in two parts. First, the generation of the dataset used to train ML algorithms is explained. Then, the ML algorithms themselves are described as well as their performances. 

	\subsection{Dataset creation}

Nowadays, the main concern when working with ML algorithm is the generation of a clean dataset rather than the algorithm itself. Why so ? Because it exists many libraries that already contains codes for all the commonly used ML algorithms. In Python, some of the most popular libraries are TensorFlow \cite{tensorflow2015}, PyTorch \cite{pytorch2019} and Skitlearn \cite{scikitlearn2011}.\\

In the present works, the Matlab statistics and ML toolbox \cite{TheMathWorks2018} is used. All the data used for the anomaly detection problem are simulated data. Nevertheless, white noise has been added to each input and output of the model. The amplitude of these noises has been defined such that the simulated data looks like the real measurements. Furthermore, slow fluctuations have been added to the input and output boundaries pressures in order to mimic the real operating conditions. To a naked eye, it's almost impossible to differentiate simulated data from measured data.\\ 

Before generating a dataset for valve anomaly detection, it is required to model the deadband problem on the output valve. In our case, the deadband has been set to random values between 1\% and 4\% to generate different test cases.\\
The following signals have been recorded:
\begin{itemize}
	\item phase separator pressure
	\item phase separator liquid level
	\item input and output valves command
\end{itemize}

\begin{figure*}[!t]
	\centering
	\psfragfig*[width=\textwidth]{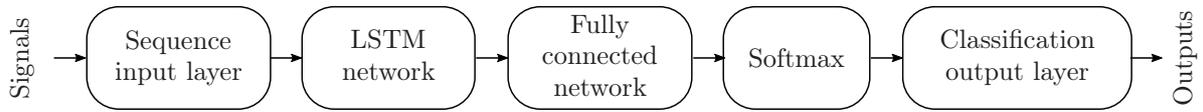}{
		\psfrag{tag0}[cc][cc][1][90]{Signals}
		\psfrag{tag1}[cc][cc][1][0]{\begin{tabular}{@{}c@{}}
   						Sequence\\
   						input layer
					  \end{tabular}}
		\psfrag{tag2}[cc][cc][1][0]{\begin{tabular}{@{}c@{}}
   						LSTM\\
   						network\\
					  \end{tabular}}
	    \psfrag{tag3}[cc][cc][1][0]{\begin{tabular}{@{}c@{}}
   						Fully\\
   						connected\\
   						network
					  \end{tabular}}
		\psfrag{tag4}[cc][cc][1][0]{Softmax}
		\psfrag{tag5}[cc][cc][1][0]{\begin{tabular}{@{}c@{}}
   						Classification\\
   						output layer
					  \end{tabular}}
		\psfrag{tag6}[cc][cc][1][90]{Outputs}
	}
	\caption{Architecture used for the LSTM network.}
	\label{fig:LSTM_architecture}
\end{figure*}

Only the valve command (and not the real position) is considered. It mimics the case where valves aren't equipped with position indicator.\\
In total, 500 times series of 60 seconds have been simulated. The dataset has been perfectly balanced: in half of the cases the valve was subject to deadband and on the other half it wasn't.\\
For the two ML algorithms described in the next sections, we use a standard cross-validation method. So, the overall dataset has been decomposed in a training set (60\% of the data), a validation set (20\% of the data) and a testing set (20\% of the data). Thus we are able to perform hyper-parameters\footnote{According to Wikipedia :"An hyperparameter is a parameter whose value is set before the learning process begins. By contrast, the values of other parameters are derived via training".} tuning for each tested ML algorithms.

	\subsection{solution 1: classification learner}
		
The first solution is to use a classification learner to determine if a valve is faulty or not. Nevertheless, those kind of algorithms require features as input and not time series. So, features where extracted from each time series of the dataset. As we don't know which features would be most suitable to identify a deadband problem, we calculate all the most common one (i.e. variance, peak to peak, skewness, kurtosis,...). In our case, we define 36 features which is few enough not to be concerned with limitation due to computer performances. But if it was the case, it would still be possible to use the same brute force approach and apply a principal component analysis \cite{Pearson1901,Hotelling1933} to reduce the number of features. Consequently, for all the time series of the dataset, each of the 4 measured signals has been transformed into a list of 9 features which could be used as input for a classification learner.\\

Once again, as there is no methodology to choose the best classification algorithm, we train multiple ones and selected the one with the highest accuracy. Thanks to parallel computing it takes less than a few minutes to train multiple algorithms including decision trees, support vector machine (SVM), logistic regression and nearest neighbors.\\
It appears that SVM with gaussian kernel \cite{Cortes1995,cristianini2000} gets the best performances among the other algorithms. SVM with gaussian kernel is particularly suitable for our problem as we have a small number of features (less than 1000) and not too much data to get concerned with computation time issue.\\

The final results obtained with the SVM are given in table \ref{tab:prediction_result}. They are compared to the results obtained with another method: a deep neural network presented in the next section.

	\subsection{solution 2: deep network}

The second idea while developing the valve anomaly detection consists in a Long Short Therm Memory (LSTM) network \cite{Hochreiter1997}. The main advantage of this deep learning algorithm is the fact that time series signals could be directly used as network inputs. It means that there is no need to calculate features in that case. Nevertheless, it generally requires more data to train this kind of network than for an SVM.\\

The architecture of the LSTM network is given in Figure \ref{fig:LSTM_architecture}. As one can see, the network is decomposed in five layers:
\begin{itemize}
	\item{The sequence input layer used to sequence data to the network}
	\item{The LSTM layer that learns long-term dependencies between time steps in sequence data}
	\item{The fully connected layer that applies weight and bias to the LSTM output in order to predict the right label}
	\item{The softmax layer that applies a softmax function to calculate the probability associated to each case (in our case normal operation of deadband problem)}
	\item{The classification output layer that provides the final prediction depending on the probability calculated in the previous layer}
\end{itemize}
			
In total, it took 200 training epochs with a constant learning rate of 0.001 to train the network. This took less than 1 minute of computation time.
			
	\subsection{Prediction results} \label{sec:Prediction results}
		
In this section, we compare the performances of the synthesized SVM and the LSTM algorithm. The comparison is based upon usual ML metrics : accuracy, precision, recall and F1 score \footnote{weighted average of the precision and recall}. More details about those metrics is available in \cite{Shung2015}. Comparison is done on a test set of 100 time series used only for this purpose (and not for training). Results are given in Table \ref{tab:prediction_result}.

\begin{table}[!htb]
	\centering
	\begin{tabular}{|c|c|c|}
		\hline
		Metrics     & SVM   & LSTM\\
		\hline
		Accuracy    & 0.98 & 0.93 \\
		\hline
		Precision   & 0.97 & 0.88 \\
		\hline	
		Recall      & 0.99 & 0.98 \\
		\hline
		$F_1$ score & 0.98 & 0.93 \\
		\hline
	\end{tabular}
	\caption{Performance indexes comparison between SVN and LSTM}
	\label{tab:prediction_result}
\end{table}

As one can see, both SVM and LSTM algorithms show good results in terms of error predictions. Nevertheless, the SVM results are slightly better. As the implementation complexity of those two algorithms is quite similar, the best option would be to deploy a SVM algorithm on the system to get an online anomaly detector. It's worth mentioning that the anomaly detection has been tested on the cryomodules only to remain consistent with the rest of the article. Nevertheless, it would be more interesting to generate an anomaly detector for process critical elements such as rotating machines of the cryogenic system: the turbines and the compressors.

\section{Conclusion}

Cryomodule models based on the first principle equation are useful to reduce downtime and to improve system performances. The first presented application of the model is the use of a linear quadratic regulator to improve the pressure stability in the phase separator. The model application second is the synthesis of a virtual sensor that predicts the RF losses in the resonator which is an indicator of the cavity thermal state. The third application is the creation of a valve anomaly detector which could help operator to detect faults in a large system.\\
Furthermore, such model could also be used for other applications like operator training and system energy cost optimization. Thus, a model should be seen as the first stone for many applications that could reduce operating cost by preventing downtime or by the mean of process optimization.

\printbibliography

\end{document}